\documentclass[%
 reprint,
superscriptaddress,
amsmath,amssymb,
aps,
pra,
]{revtex4-1}
\usepackage{caption}
\usepackage{subcaption}
\usepackage{graphicx}
\usepackage{xcolor}
\usepackage{graphicx}
\usepackage{dcolumn}
\usepackage{bm}

\begin{document}

\preprint{APS/123-QED}

\title{Space--Time Surface Plasmon Polaritons: A New Propagation-Invariant Surface Wave Packet}

\author{Kenneth L. Schepler}
\affiliation{CREOL, The College of Optics \& Photonics, University of Central Florida, Orlando, FL 32816, USA}
\author{Murat Yessenov}
\affiliation{CREOL, The College of Optics \& Photonics, University of Central Florida, Orlando, FL 32816, USA}
\author{Yertay Zhiyenbayev}
\altaffiliation{Current address: Nazarbayev University, Nur-Sultan,
Kazakhstan}
\author{Ayman F. Abouraddy}
 \email{Corresponding author: raddy@creol.ucf.edu}
\affiliation{CREOL, The College of Optics \& Photonics, University of Central Florida, Orlando, FL 32816, USA}




\begin{abstract}
We introduce the unique class of propagation-invariant surface plasmon polaritons (SPPs) representing pulsed surface wave packets propagating along unpatterned metal-dielectric interfaces and are localized in all dimensions -- with potentially subwavelength transverse spatial widths. The characteristic features of such linear diffraction-free, dispersion-free `plasmonic bullets' stem from tight spatio-temporal correlations incorporated into the SPP spectral support domain, and we thus call them `space-time' SPPs. We show that the group velocity of space-time SPP wave packets can be readily tuned to subluminal, superluminal, and even negative values by tailoring the spatio-temporal field structure independently of any material properties. We present an analytical framework and numerical simulations for the propagation of space-time SPPs in comparison with traditional pulsed SPPs whose spatial and temporal degrees of freedom are separable, thereby verifying the propagation-invariance of the former.
\end{abstract}

\maketitle

\section{Introduction}

\begin{figure}[t!]
  \begin{center}
  \includegraphics[width=8.6cm]{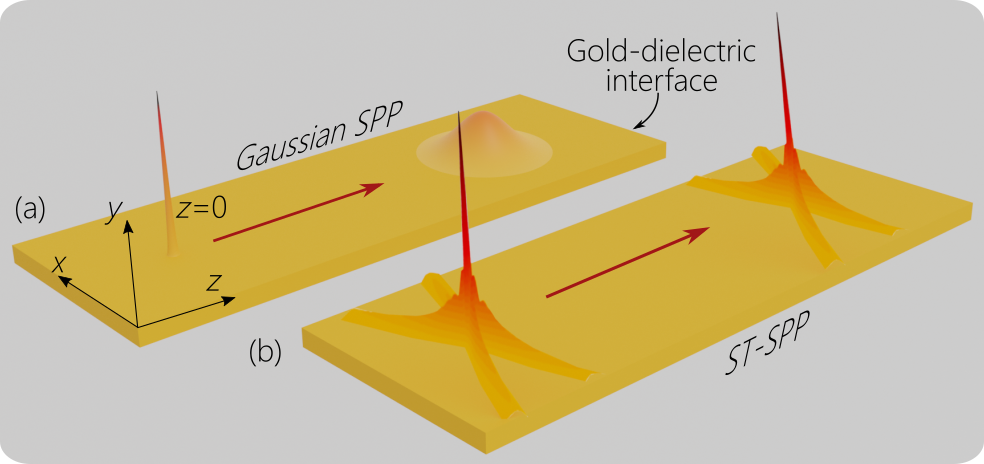}
  \end{center}
  \caption{Conceptual depiction of the propagation of (a) a traditional Gaussian SPP wave packet in comparison with (b) a ST-SPP along a metal-dielectric interface. Both are pulsed wave packets that are localized along $z$, are strongly confined at the interface along $y$, and are initially localized along the transverse dimension $x$. We plot the inital spatio-temporal intensity $I(x,y\!=\!0,z\!=\!0;t)$ and after propagation along $z$ (the field is confined to the interface along $y$). The traditional SPP in (a) undergoes diffractive and dispersive spreading, whereas the ST-SPP in (b) is immune to both.}
  \label{STSPP_concept}
\end{figure}

Surface plasmon polaritons (SPPs) are surface waves confined to the interface between a dielectric and a metal below its plasmon resonance frequency \cite{Otto68ZP,Economou69PR,Kretschman71ZP,Raether88Book,Maier07Book,Zhang2012}. However, the spatial profile of the SPP undergoes diffractive spreading along the unbounded transverse dimension upon free propagation, which mars the prospect of strongly confining SPPs in all dimensions. Moreover, SPPs are intrinsically dispersive, such that the temporal width of a broadband SPP pulse increases with propagation. The combined effect on a Gaussian SPP wave packet of diffractive and dispersive spreading in space and time is highlighted in Fig.~\ref{STSPP_concept}a. We aim here to propose a new \textit{propagation-invariant} SPP wave packet that is localized in \textit{all} dimensions. 

A potential avenue to combat SPP diffractive spreading is to rely on particularly shaped profiles corresponding to so-called `diffraction-free' beams, such as Bessel \cite{Durnin87PRL}, Mathieu \cite{Gutierrez2003AJP}, or Weber \cite{Bandres04OL} beams. However, none of these are useful for producing diffraction-free SPPs because they fundamentally require for their realization \textit{two} transverse dimensions \cite{Levy16PO}. For example, an optical field whose distribution conforms to a Bessel function in one dimension diffracts, in contrast to its two-dimensional counterpart that does not. Indeed, it is now understood that the dimensionality of the transverse profile of an optical field plays a critical role in the prospect of achieving diffraction-free behavior. Self-similar SPP propagation requires that a one-dimensional (1D) diffraction-free spatial profile be identified beyond the trivial cases of plane or cosine waves \cite{Lin12PRL} that are not localized. However Berry proved that there are \textit{no} 1D diffraction-free optical beams, except for 1D Airy beams that travel along a curved trajectory rather than a straight line \cite{Berry79AMP,Unnikrishnan96AMP,Siviloglou07OL,Siviloglou07PRL,Hu2012,Efremidis19Optica}. This beam profile has been recently proposed for plasmonic platforms and called `Airy plasmons' \cite{Salandrino10OL}, which have been realized in a variety of settings \cite{Minovich2011,Zhang11OL,Li2011PRL,Wang2017OpEx,Yin2018OpEx,Minovich14LPR}. The current state-of-the-art therefore suggests that -- fundamentally -- no avenues exist for producing localized, diffraction-free SPPs that travel in a \textit{straight line} because of the unavoidable restrictions imposed by the reduced-dimensionality intrinsic to plasmonic surface waves. 

Berry's formulation, however, presumed \textit{monochromatic} fields \cite{Berry79AMP,Unnikrishnan96AMP,Siviloglou07OL}. If one relaxes the monochromaticity constraint by considering finite-bandwidth optical \textit{pulses}, it can be shown that 1D \textit{propagation-invariant} pulsed beams (or wave packets) of arbitrary profile that travel in a straight line can indeed be synthesized \cite{Dallaire09OE,Jedrkiewicz13OE,Kondakci16OE}. By structuring the spatio-temporal spectrum of the field such that each spatial frequency is precisely associated with one wavelength \cite{Donnelly93PRSLA,Saari04PRE,Longhi04OE}, the 1D wave packet is rendered diffraction-free \textit{and} dispersion-free independently of the characteristics of the medium. Such `space-time' (ST) wave packets have been recently synthesized in free space \cite{Kondakci17NP,Yessenov19OPN} and a host of their intriguing features have been observed, such as self-healing \cite{Kondakci18OL}, controllable group velocities in free space \cite{Kondakci19NC} and in dielectrics \cite{Bhaduri19Optica}, long-distance propagation \cite{Bhaduri18OE,Bhaduri19OL}, omni-resonance in planar cavities \cite{Shiri19arxiv}, and anomalous refraction at dielectric interfaces \cite{Bhaduri19unpublished}. Indeed, using this methodology, ST Airy wave packets have been observed that travel in straight lines, and accelerate instead in space-time \cite{Kondakci18PRL}. These ST light sheets are 1D versions of previously studied localized wave packets \cite{Reivelt03arxiv,Kiselev07OS,Turunen10PO,FigueroaBook14}.

Here we propose a new class of surface wave packets for plasmonic platforms that we call `space-time' surface plasmon polaritons (ST-SPPs). They are confined in the direction normal to the metal-dielectric interface in the same manner as traditional SPPs, but are diffraction-free in the unbounded transverse dimension by virtue of their intrinsic spatio-temporal structure \cite{Kondakci16OE,Parker16OE,Kondakci17NP,Porras17OL,Efremidis17OL,Wong17ACSP1,Wong17ACSP2,PorrasPRA18,Kondakci19ACSP}. Furthermore, despite being pulsed solutions, they are nevertheless inherently immune to dispersive spreading \cite{Sonajalg96OL,Sonajalg97OL}. These surface-confined propagation-invariant pulsed SPPs are localized in \textit{all} dimensions and are transported rigidly along the metal-dielectric interface, thereby constituting linear `plasmonic bullets'. In principle, ST-SPPs may take on an arbitrary spatio-temporal profile, and may assume any transverse spatial extent and pulse width. Although \textit{free-space} ST wave packets can\textit{not} achieve subwavelength transverse widths, such dimensions \textit{can} indeed be realized in ST-SPPs because of the intrinsic dispersion of the SPP light-cone in comparison to that of free space. Unlike Airy plasmons, ST-SPPs travel in a straight line while retaining similar self-healing characteristics. Furthermore, the group velocity of ST-SPPs can assume arbitrary values, above (superluminal) or below (subluminal) that of a traditional SPP, or even negative values (backward propagation) by appropriately varing the field spatio-temporal structure. We show through analysis and simulations that ST-SPPs circumvent the two deleterious effects highlighted in Fig.~\ref{STSPP_concept}a that are unavoidably associated with traditional SPPs: diffractive spreading in the transverse dimension, and axial dispersive broadening of ultrashort pulsed SPPs. In contrast, as shown in Fig.~\ref{STSPP_concept}b, both these effects are neutralized when utilizing ST-SPPs. These results may help unlock the latent potential of plasmonics in its simplest structural embodiment at a metal-dielectric interface, including the realization of large field enhancements and the potential for backward phase-matching \cite{Lan15NM}.

The paper is organized as follows. First, we review the formulation of ST wave packets in unbounded media and that of finite-transverse-width SPPs. We then introduce the theory of ST-SPPs that combines these two disparate concepts -- emphasizing the spectral representation of ST-SPPs on the SPP light-cone and their classification according to group velocity. Next, we present simulations contrasting the propagation of SPPs and ST-SPPs launched with the same initial spatial and temporal widths (including subwavelength spatial widths), verifying the propagation-invariance of ST-SPPs. After commenting on the impact of plasmonic losses and dispersion, we propose a pathway to synthesizing ST-SPPs and suggest avenues for further investigations before presenting our conclusions.

\section{Space-time wave packets}

By expanding a pulsed optical field propagating in a dielectric of refractive index $n$ into monochromatic plane waves $e^{i(k_{x}x+k_{y}y+k_{z}z-\omega t)}$, the envelope $\psi(x,y,z;t)$ of the wave packet $E(x,y,z;t)\!=\!e^{i(nk_{\mathrm{o}}z-\omega_{\mathrm{o}}t)}\psi(x,y,z;t)$ can be expressed as follows \cite{SalehBook07}:
\begin{equation}\label{Eq:GenWavePkt}
\begin{aligned}
&\psi(x,y,z;t)\!=\\
&\!\!\iiint\!\!dk_{x}dk_{y}d\Omega\,\widetilde{\psi}(k_{x},k_{y},\Omega)e^{i(k_{x}x+k_{y}y+(k_{z}-nk_{\mathrm{o}})z-\Omega t)} \; ; 
\end{aligned}
\end{equation}
where $k_{x}$, $k_{y}$, and $k_{z}$ are the components of the wave vector (assumed all to be real) along the Cartesian coordinates $x$, $y$, and $z$ (the propagation direction), $\omega$ is the temporal (angular) frequency, the spatio-temporal spectrum $\widetilde{\psi}(k_{x},k_{y},\Omega)$ is the three-dimensional Fourier transform of $\psi(x,y,0;t)$, $\omega_{\mathrm{o}}$ is the carrier frequency, $k_{\mathrm{o}}\!=\!\omega_{\mathrm{o}}/c$ the corresponding wave number, $c$ is the speed of light in vacuum, and $\Omega\!=\!\omega-\omega_{\mathrm{o}}$ is the frequency measured with respect to $\omega_{\mathrm{o}}$. The wave-vector components satisfy the dispersion relationship $k_{x}^{2}+k_{y}^{2}+k_{z}^{2}\!=\!n^{2}(\tfrac{\omega}{c})^{2}$, which corresponds geometrically to the surface of a hyper-cone (referred to hereon as the `light-cone'). A \textit{monochromatic} plane wave corresponds to a point on the surface of the light-cone, so that the spatio-temporal spectral support for a traditional \textit{pulsed beam} is represented by an extended domain on the surface of the light-cone.

Propagation-invariant ST wave packets have a constraint imposed on the \textit{support domain} of their spatio-temporal spectrum \cite{Donnelly93PRSLA}: $\Omega\!=\!(k_{z}-nk_{\mathrm{o}})\:c\tan{\theta}$, which corresponds to a hyperplane $\mathcal{P}(\theta)$ tilted an angle $\theta$ with respect to the $k_{z}$-axis. The ST wave packet is thus confined to the intersection of the light-cone with $\mathcal{P}(\theta)$. The envelope of the ST wave packet becomes
\begin{equation}\label{Eq:FreeSTWavePacket}
\begin{aligned}
&\psi(x,y,z;t)\!=\\
&\!\!\iint\!\!dk_{x}dk_{y}\widetilde{\psi}(k_{x},k_{y})e^{i(k_{x}x+k_{y}y)}e^{-i\Omega(t-z/\widetilde{v})}=\psi(x,y,0;t-z/\widetilde{v}),
\end{aligned}
\end{equation}
such that the wave packet is now propagation-invariant, traveling rigidly at a group velocity $\widetilde{v}\!=\!c\tan{\theta}$ (group index $\widetilde{n}\!=\!\cot{\theta}$) that is \textit{independent} of the refractive index and depends only on $\theta$, which we denote the spectral tilt angle. Note that $\theta$ is \textit{not} a physical angle, but a mathematical description of the spatio-temporal spectral correlations introduced into the ST wave packet \cite{Bhaduri18OE,Yessenov19PRA,Yessenov19OE}. The reduced-dimensionality spectrum $\widetilde{\psi}(k_{x},k_{y})$ is the two-dimensional Fourier transform of $\psi(x,y,0;0)$. The frequency $\Omega$ in Eq.~\ref{Eq:FreeSTWavePacket} is no longer an independent variable as it is in Eq.~\ref{Eq:GenWavePkt}; rather, $\Omega\!=\!\Omega(k_{x},k_{y};\theta)$ is determined by the hypercurve representing the intersection of $\mathcal{P}(\theta)$ with the light-cone.

A crucial aspect of ST wave packets for the development of ST-SPPs can be seen in Eq.~\ref{Eq:FreeSTWavePacket}: the propagation-invariance of the ST wave packet \textit{is independent of dimensionality} \cite{Kondakci16OE}, in contradistinction to traditional monochromatic diffraction-free beams that require two transverse dimensions for their realization \cite{Levy16PO}. Consequently, 1D ST wave packets can be constructed by holding the field uniform along $y$ (that is, $k_{y}\!=\!0$) and dropping the integral over $k_{y}$ in Eq.~\ref{Eq:FreeSTWavePacket}; $\widetilde{\psi}(k_{x},k_{y})\!\rightarrow\!\widetilde{\psi}(k_{x})$. In this case, the dispersion relationship $k_{x}^{2}+k_{z}^{2}\!=\!n^{2}(\tfrac{\omega}{c})^{2}$ corresponds to a light-cone in three-dimensional $(k_{x},k_{z},\tfrac{\omega}{c})$-space, and $\mathcal{P}(\theta)$ becomes a plane, rather than a hyperplane, that intersects with the light-cone in a conic section. The `luminal' condition $\widetilde{n}\!=\!n$, or $\widetilde{v}\!=\!c/n$, corresponds to a spectral plane $\mathcal{P}(\arctan{\tfrac{1}{n}})$ that is tangential to the light-cone. The subluminal condition $\widetilde{n}\!>\!n$, or $\widetilde{v}\!<\!c/n$, corresponds to $\mathcal{P}(\theta\!<\arctan{\tfrac{1}{n}})$ intersecting with the light-cone in an ellipse. The superluminal condition $\widetilde{n}\!<\!n$, or $\widetilde{v}\!>\!c/n$, corresponds to $\mathcal{P}(\theta\!>\!\arctan{\tfrac{1}{n}})$ intersecting with the light-cone in a hyperbola, a parabola, or an ellipse, depending on the value of $\theta$ \cite{Yessenov19PRA}. Negative group velocities $\widetilde{v}\!<\!0$ are obtained when $\theta\!>\!90^{\circ}$, corresponding to backward propagation (without violating relativistic causality \cite{Shaarawi00JPA,SaariPRA18,Yessenov19PRA,Saari19PRA,Yessenov19OE,Saari20PRA}). These features have all been confirmed experimentally in free space \cite{Kondakci19NC,Yessenov19OE} and dielectrics \cite{Bhaduri19Optica,Bhaduri19unpublished}. We demonstrate below that similar control can be exercised on ST-SPPs.

\section{Theory of ST-SPP wave packets}

\subsection{Traditional SPP wave packets of finite transverse spatial width}

\begin{figure*}[t!]
  \begin{center}
  \includegraphics[width=14cm]{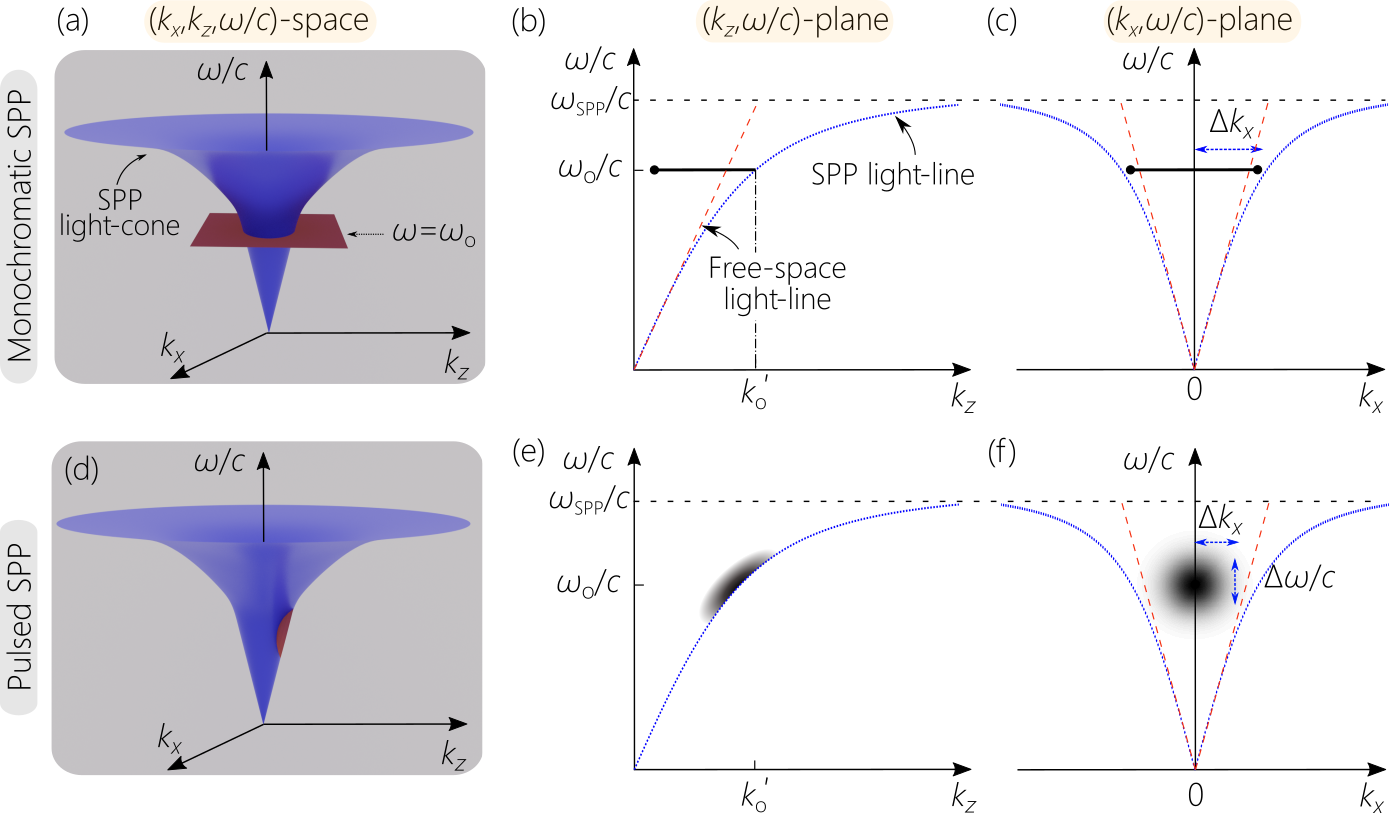}
  \end{center}
  \caption{Monochromatic and pulsed SPPs of \textit{finite} transverse spatial extent and their spectral representation on the SPP light-cone in $(k_{x},k_{z},\tfrac{\omega}{c})$-space. (a) The spectral support domain for a \textit{monochromatic} SPP lies along the intersection of the SPP light-cone with a horizontal iso-frequency plane $\omega\!=\omega_{\mathrm{o}}$. (b) Spectral projection for the monochromatic SPP onto the $(k_{z},\tfrac{\omega}{c})$-plane and (c) onto the $(k_{x},\tfrac{\omega}{c})$-plane. (d) Spectral support domain for a \textit{pulsed} SPP on the SPP light-cone, and its projection (e) onto the $(k_{z},\tfrac{\omega}{c})$-plane and (f) onto the $(k_{x},\tfrac{\omega}{c})$-plane, depicted as 2D domains. The spatial and temporal spectral degrees of freedom are separable; $\Delta k_{x}$ is the spatial bandwidth and $\Delta\omega$ is the temporal bandwidth. The thin dotted blue curves are the SPP light-lines, the dashed red lines are the dielectric light-lines, and the thick black solid lines are the projected SPP spectral support domains.}
  \label{Fig:SPPs}
\end{figure*}

We briefly review the formulation of traditional SPPs of finite transverse spatial width to set the stage for ST-SPPs. The coordinate system for SPPs is shown in Fig.~\ref{STSPP_concept}a, where $z$ is the propagation direction along the metal-dielectric interface, $y$ is the direction of confinement normal to the interface, and $x$ is the transverse coordinate defining the SPP width. Implementing the boundary conditions at the interface for TM polarization yields the SPP dispersion relationship $k_{x}^{2}+k_{z}^{2}\!=\!(\tfrac{\omega}{c})^{2}\tfrac{\epsilon_{\mathrm{m}}(\omega)\epsilon_{\mathrm{d}}}{\epsilon_{\mathrm{m}}(\omega)+\epsilon_{\mathrm{d}}}$, which corresponds geometrically to a modified SPP light-cone (Fig.~\ref{Fig:SPPs}a); here $\epsilon_{\mathrm{d}}$ and $\epsilon_{\mathrm{m}}$ are the relative permittivities of the dielectric and metal, respectively. We make use of the Drude model for the metal permittivity, the real part of which is $\epsilon_{\mathrm{m}}(\omega)=1-\frac{\omega_{\mathrm{p}}^{2}}{\omega^{2}+\Gamma^{2}}$, so that the SPP light-cone becomes
\begin{equation}\label{Eq:SPP_LightCone}
k_{x}^{2}+k_{z}^{2}=\left(\frac{\omega}{c}\right)^{2}\frac{\epsilon_{\mathrm{d}}}{1+\epsilon_{\mathrm{d}}}\,\frac{\omega^{2}-\omega_{\mathrm{p}}^{2}+\Gamma^{2}}{\omega^{2}-\omega_{\mathrm{p}}'^{2}+\Gamma^{2}},
\end{equation}
where $\omega_{\mathrm{p}}$ is the metal plasmon resonance frequency, $\Gamma$ is the damping rate, $\omega_{\mathrm{p}}'\!=\!\omega_{\mathrm{p}}/\sqrt{1+\epsilon_{\mathrm{d}}}$, $\omega\!<\!\omega_{\mathrm{SPP}}$, where $\omega_{\mathrm{SPP}}^{2}\!=\!\omega_{\mathrm{p}}'^{2}-\Gamma^{2}$ is the maximum admissible frequency for the SPP, and the curved light-line (when $k_{x}\!=\!0$) indicates that the SPP is dispersive. Throughout, we make use of the parameters for gold in Table~\ref{tbl:input} \cite{Olmon}.

The spectral support domain of a \textit{monochromatic} SPP is the circle at the intersection of the SPP light-cone with a horizontal iso-frequency plane $\omega\!=\!\omega_{\mathrm{o}}$ (Fig.~\ref{Fig:SPPs}a-c). Such a SPP undergoes diffractive spreading. A \textit{pulsed} SPP of finite spatial and temporal extents is represented by a 2D spectral support domain on the surface of the SPP light-cone (Fig.~\ref{Fig:SPPs}d-f). The spatio-temporal spectrum (Fig.~\ref{Fig:SPPs}f) is typically separable with respect to the spatial and temporal degrees of freedom, $k_{x}$ and $\omega$, respectively, and the SPP thereby undergoes both diffractive and dispersive spreading.

\subsection{ST-SPP wave packets}
Unlike the 2D spectral support domain for a pulsed SPP wave packets (Fig.~\ref{Fig:SPPs}d-f), the support domain for a ST-SPP has a reduced dimensionality corresponding to the intersection of the SPP light-cone with a tilted spectral plane $\mathcal{P}(\theta)$ described by the equation $\Omega\!=\!(k_{z}-k_{\mathrm{o}}')c\tan{\theta}$ (Fig.~\ref{Fig:SpectralSupportSubSuper}a). The wave number $k_{\mathrm{o}}'$ corresponds to the frequency $\omega_{\mathrm{o}}$and lies on the SPP light-line ($k_{\mathrm{o}}'\!\neq\!\tfrac{\omega_{\mathrm{o}}}{c}$). We exclude negative axial wave numbers $k_{z}\!<\!0$ because they violate causal excitation and propagation \cite{Shaarawi95JMP}, and include only points \textit{on} the surface of the SPP light-cone (others correspond to evanescent waves) \cite{Yessenov19PRA}. Consequently, the spectral projection onto the $(k_{z},\tfrac{\omega}{c})$-plane is restricted to a \textit{straight line} that is tilted by an angle $\theta$ with respect to the $k_{z}$-axis, as in the case of freely propagating ST wave packets \cite{Yessenov19PRA}. Within this framework, the frequency $\omega_{\mathrm{o}}$ occurs at $k_{x}\!=\!0$ on the SPP light-line and corresponds either to the maximum \textit{or} the minimum frequency in the pulse spectrum (see below).  

The envelope for the electric and magnetic field components of the ST-SPP in the dielectric are thus given by:
\begin{widetext}
\begin{eqnarray}\label{Eq:STSPPInSpaceTimeInDielectric}
\psi_{x}^{(\mathrm{d})}(x,y,z;t)&=&\int\!\!dk_{x}\,\widetilde{E}_{\mathrm{o}}(k_{x})\frac{k_{x}}{k_{z}}e^{i(k_{x}x+k_{y_{\mathrm{d}}}y)}e^{-i\Omega(k_{x};\theta)(t-z/\widetilde{v})},\\
\psi_{y}^{(\mathrm{d})}(x,y,z;t)&=&-\int\!\!dk_{x}\,\widetilde{E}_{\mathrm{o}}(k_{x})\frac{k_{x}^{2}+k_{z}^{2}}{k_{z}k_{y_{\mathrm{d}}}}e^{i(k_{x}x+k_{y_{\mathrm{d}}}y)}e^{-i\Omega(k_{x};\theta)(t-z/\widetilde{v})},\\
\psi_{z}^{(\mathrm{d})}(x,y,z;t)&=&\int\!\!dk_{x}\,\widetilde{E}_{\mathrm{o}}(k_{x})e^{i(k_{x}x+k_{y_{\mathrm{d}}}y)}e^{-i\Omega(k_{x};\theta)(t-z/\widetilde{v})},\\
\xi_{x}^{(\mathrm{d})}(x,y,z;t)&=&\int\!\!dk_{x}\,\widetilde{H}_{\mathrm{o}}(k_{x})e^{i(k_{x}x+k_{y_{\mathrm{d}}}y)}e^{-i\Omega(k_{x};\theta)(t-z/\widetilde{v})},\\
\xi_{y}^{(\mathrm{d})}(x,y,z;t)&=&0,\\
\xi_{z}^{(\mathrm{d})}(x,y,z;t)&=&-\int\!\!dk_{x}\,\widetilde{H}_{\mathrm{o}}(k_{x})\frac{k_{x}}{k_{z}}e^{i(k_{x}x+k_{y_{\mathrm{d}}}y)}e^{-i\Omega(k_{x};\theta)(t-z/\widetilde{v})};
\end{eqnarray}
\end{widetext}

where $\psi_{j}^{(\mathrm{d})}(x,y,z;t)$ is the envelope of the electric field component $E_{j}^{(\mathrm{d})}(x,y,z;t)$, $\xi_{j}^{(\mathrm{d})}(x,y,z;t)$ is the envelope of the magnetic field component $H_{j}^{(\mathrm{d})}(x,y,z;t)$, $j\!=\!x,y,z$, $\widetilde{E}_{\mathrm{o}}(k_{x})$ is the spatial spectrum of the ST-SPP electric-field components, $\widetilde{H}_{\mathrm{o}}(k_{x})$ is the corresponding spatial spectrum for the magnetic field components, $\widetilde{E}_{\mathrm{o}}\!=\!\tfrac{\eta_{\mathrm{o}}}{\sqrt{\epsilon_{\mathrm{d}}+\epsilon_{\mathrm{m}}(\omega)}}\widetilde{H}_{\mathrm{o}}$, $\eta_{\mathrm{o}}$ is the impedance of free space, $k_{y_{\mathrm{d}}}$ is the $y$-component of the wave vector in the dielectric and is given by the usual expression $k_{y_{\mathrm{d}}}\!=\!\tfrac{\omega}{c}\tfrac{\epsilon_{\mathrm{d}}}{\sqrt{\epsilon_{\mathrm{d}}+\epsilon_{\mathrm{m}}(\omega)}}$. The equations for the field components in the metal take on a similar form with $k_{y_{\mathrm{d}}}$ replaced by $k_{y_{\mathrm{m}}}\!=\!\tfrac{\omega}{c}\tfrac{\epsilon_{\mathrm{m}}(\omega)}{\sqrt{\epsilon_{\mathrm{d}}+\epsilon_{\mathrm{m}}(\omega)}}$. We take $y\!>\!0$ in the dielectric and $y\!<\!0$ in the metal ($k_{y_{\mathrm{d}}}$ is a negative imaginary number and $k_{y_{\mathrm{m}}}$ positive imaginary). 

Instead of the spatial and temporal frequencies $k_{x}$ and $\Omega$, respectively, being independent variables as in Eq.~\ref{Eq:GenWavePkt} and Fig.~\ref{Fig:SPPs}, they are now \textit{dependent} variables, $\Omega\!=\!\Omega(k_{x};\theta)$. After eliminating $k_{z}$ from the SPP dispersion relationship in Eq.~\ref{Eq:SPP_LightCone} by substituting the equation for the spectral plane $\mathcal{P}(\theta)$, we obtain the relationship $\Omega(k_{x};\theta)$ that describes the spectral support domain for the ST-SPP:
\begin{equation}\label{Eq:kx_omega_ST-SPP}
k_{x}^{2}=\left(\frac{\omega}{c}\right)^{2}\frac{\epsilon_{\mathrm{d}}}{1+\epsilon_{\mathrm{d}}}\,\frac{\omega^{2}-\omega_{\mathrm{p}}^{2}+\Gamma^{2}}{\omega^{2}-\omega_{\mathrm{p}}'^{2}+\Gamma^{2}}-\left(k_{\mathrm{o}}'+\frac{\omega-\omega_{\mathrm{o}}}{c}\cot{\theta}\right)^{2},
\end{equation}
which differs significantly from its free-space counterpart that is a standard conic section \cite{Yessenov19PRA}. We evaluate $\Omega(k_{x};\theta)$ numerically, which is then the basis for the synthesis of the ST-SPP \cite{Kondakci17NP,Yessenov19OPN}.

In general, this formulation is a spatio-temporal extension of Fourier optics implemented in the plasmonics context. The only independent variables are $k_{x}$ and $\theta$ that together dictate $\Omega(k_{x};\theta)$, which in turn determines $k_{z}$, $k_{y_{\mathrm{d}}}$, and $k_{y_{\mathrm{m}}}$. Although the ST-SPP has a finite spatial bandwdith and may have a narrow transverse profile, it nevertheless propagates at the metal-dielectric interface self-similarly because all the electric and magnetic field components, in both the dielectric \textit{and} the metal, now take the general form:
\begin{equation}
\psi_{x}^{(\mathrm{d})}(x,y,z;t)=\psi_{x}^{(\mathrm{d})}(x,y,0;t-z/\widetilde{v}).
\end{equation}
The ST-SPP therefore propagates rigidly at a group velocity $\widetilde{v}\!=\!\tfrac{\partial\omega}{\partial k_{z}}\!=\!c\tan{\theta}$ (group index $\widetilde{n}\!=\!\cot{\theta}$), and the ST-SPP is thus free of dispersion and diffraction (Fig.~\ref{STSPP_concept}b). It is critical to appreciate that the \textit{internal} degree of freedom $\theta$ of the ST-SPP determines the group velocity \textit{independently} of the material properties and the field profile. The spatial profile at the pulse center is $\psi_{x}^{(\mathrm{d})}(x,0,0;0)\!=\!\int\!dk_{x}\,\widetilde{E}_{\mathrm{o}}(k_{x})\tfrac{k_{x}}{k_{z}}e^{ik_{x}x}$, and the temporal profile at the ST-SPP center is  $\widetilde{E}_{\mathrm{o}}(k_{x})$.  $\psi_{x}^{(\mathrm{d})}(0,0,0;t)\!=\!\int\,dk_{x}\widetilde{E}_{\mathrm{o}}(k_{x})\tfrac{k_{x}}{k_{z}}e^{-i\Omega(k_{x};\theta)t}$, both of which are thus determined by the spatial spectrum $\widetilde{E}_{\mathrm{o}}(k_{x})$ and $\Omega(k_{x};\theta)$. The ST-SPP profile can thus take on a Gaussian rather than an Airy profile, and can even take on an Airy profile while avoiding self-acceleration \cite{Kondakci18PRL}.

\subsection{Defining subluminal and superluminal ST-SPPs}

The equations for the field components given take the same form for all values of $\theta$. In contrast to the free-space light-cone, the surface of the SPP light-cone is \textit{not} convex, so that $\mathcal{P}(\theta)$ may intersect with the SPP light-cone in two disjoint sections in some cases (as specified below). Moreover, this lack of convexity and the existence of an upper frequency limit $\omega\!<\!\omega_{\mathrm{SPP}}$ impact the delineation of the `subluminal' and `superliminal' regimes, and limit the exploitable bandwidth for ST-SPPs.

\begin{figure*}
  \begin{center}
  \includegraphics[width=7.6cm]{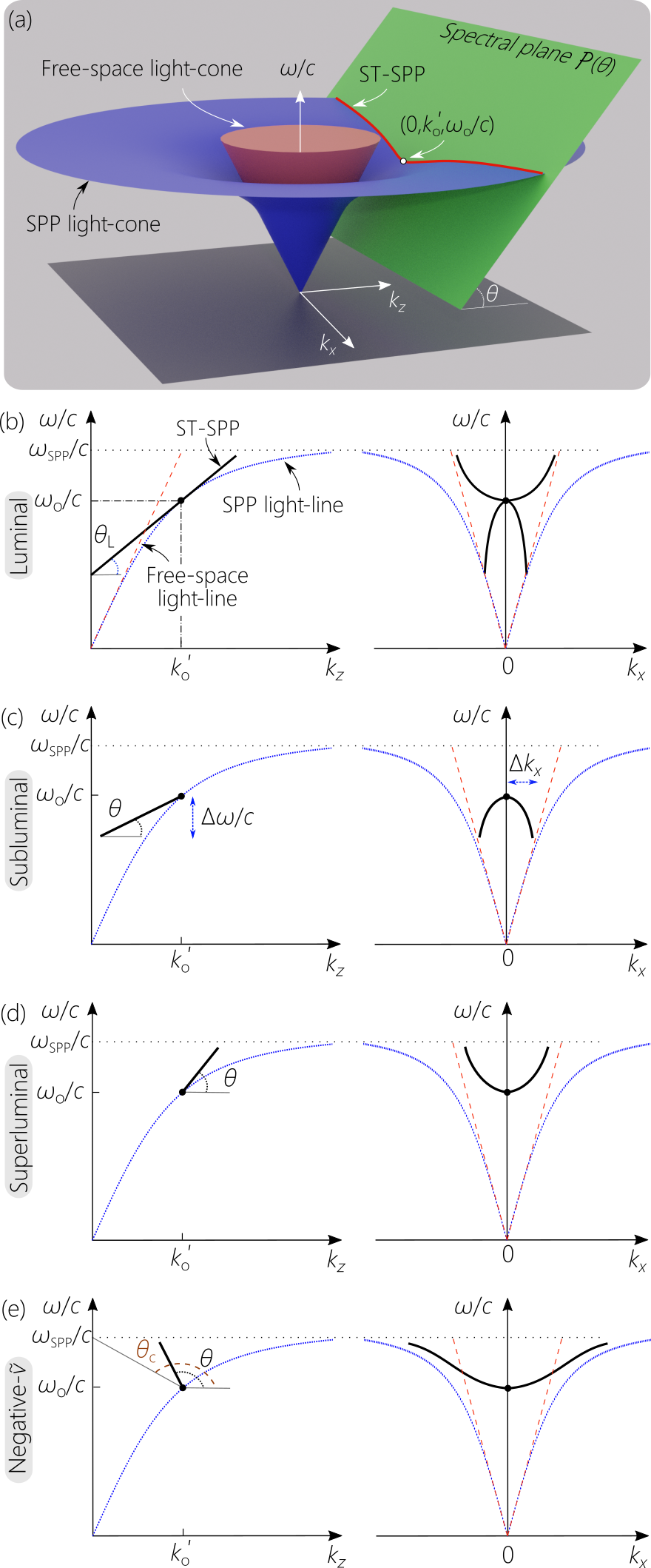}
  \end{center}
  \caption{(a) Three-dimensional spectral representation of the support domain for a ST-SPP at the intersection of the SPP light-cone with a tilted spectral plane $\mathcal{P}(\theta)$. (b) Representation of the support domain of \textit{luminal} ST-SPPs projected onto the $(k_{z},\tfrac{\omega}{c})$-plane (left) and onto the $(k_{x},\tfrac{\omega}{c})$-plane (right). (c) Same as (b) for \textit{subluminal} ST-SPPs. (d) Same as (b) for \textit{superluminal} ST-SPPs with positive-$\widetilde{v}$. (e) Same as (b) for \textit{negative}-$\widetilde{v}$ ST-SPPs. In (b-e), the thin dotted blue curves are the SPP light-lines, the thin dashed red lines are the dielectric light-lines, and the thick black curves are the projected ST-SPP spectral support domains.}
  \label{Fig:SpectralSupportSubSuper}
\end{figure*}

We take the `luminal' ST-SPP to be that whose group index $\widetilde{n}_{\mathrm{L}}$ has the same value of the group index of a traditional extended SPP $\widetilde{n}_{\mathrm{SPP}}$ at the same frequency $\omega\!=\!\omega_{\mathrm{o}}$, $\widetilde{n}_{\mathrm{L}}\!=\!\widetilde{n}_{\mathrm{SPP}}$. By first selecting a point $(k_{z},\tfrac{\omega}{c})\!=\!(k_{\mathrm{o}}',\tfrac{\omega_{\mathrm{o}}}{c})$ on the SPP light-line, the luminal ST-SPP has its spectral support in the plane $\mathcal{P}(\theta_{\mathrm{L}})$ that is tangential to the SPP light-cone at $\omega_{\mathrm{o}}$ (Fig.~\ref{Fig:SpectralSupportSubSuper}b),
\begin{equation}
\cot{\theta_{\mathrm{L}}}=\frac{ck_{\mathrm{o}}'}{\omega_{\mathrm{o}}}+\frac{\omega_{\mathrm{o}}}{ck_{\mathrm{o}}'}\left(\frac{\epsilon_{\mathrm{d}}}{1+\epsilon_{\mathrm{d}}}\,\,\frac{\omega_{\mathrm{o}}\omega_{\mathrm{p}}}{\omega_{\mathrm{o}}^{2}-\omega_{\mathrm{p}}^{2}+\Gamma^{2}}\right)^{2}.
\end{equation}
The luminal group velocity is $\widetilde{v}_{\mathrm{L}}\!=\!c\tan{\theta_{\mathrm{L}}}$, and the luminal group index is $\widetilde{n}_{\mathrm{L}}\!=\!\cot{\theta_{\mathrm{L}}}$; see Fig.~\ref{Fig:LuminalAngle}a for a plot of $\theta_{\mathrm{L}}(\omega_{\mathrm{o}})$. The spatio-temporal spectrum of the luminal ST-SPP has two branches above and below $\omega_{\mathrm{o}}$ having maximum temporal bandwidths of $\Delta\omega^{\mathrm{max}}\!=\!\omega_{\mathrm{SPP}}-\omega_{\mathrm{o}}$ and $\Delta\omega^{\mathrm{max}}\!=\!\widetilde{v}_{\mathrm{L}}k_{\mathrm{o}}'$, respectively (Fig.~\ref{Fig:SpectralSupportSubSuper}b). These values determine the minimum pulse widths attainable with luminal ST-SPPs.

The luminal condition at a given $\omega_{\mathrm{o}}$ separates the subluminal and superluminal regimes. The \textit{subluminal} range for ST-SPPs ($\widetilde{v}\!=\!c\tan{\theta}\!<\!\widetilde{v}_{\mathrm{L}}$) corresponds to spectral tilt angles in the range $0\!<\!\theta\!<\theta_{\mathrm{L}}$, $\omega_{\mathrm{o}}$ in this case is the \textit{maximum} frequency in the spectrum, and the maximum bandwidth is $\Delta\omega^{\mathrm{max}}\!=\!\widetilde{v}k_{\mathrm{o}}'$ (Fig.~\ref{Fig:SpectralSupportSubSuper}c). The \textit{superluminal} span of ST-SPPs ($\widetilde{v}\!>\!\widetilde{v}_{\mathrm{L}}$) corresponds to spectral tilt angles in the range $\theta_{\mathrm{L}}\!<\!\theta\!<\!180^{\circ}$, with positive-$\widetilde{v}$ values spanning $\theta_{\mathrm{L}}\!<\!\theta\!<\!90^{\circ}$ (Fig.~\ref{Fig:SpectralSupportSubSuper}d), and negative-$\widetilde{v}$ values spanning $90^{\circ}\!<\!\theta\!<\!180^{\circ}$ (Fig.~\ref{Fig:SpectralSupportSubSuper}e). Here $\omega_{\mathrm{o}}$ is the \textit{minimum} frequency in the ST-SPP spectrum. For superluminal positive-$\widetilde{v}$, the maximum temporal bandwidth is $\Delta\omega^{\mathrm{max}}\!=\!\omega_{\mathrm{SPP}}-\omega_{\mathrm{o}}$, which is independent of $\theta$ (Fig.~\ref{Fig:SpectralSupportSubSuper}d). There are two regimes for superluminal negative-$\widetilde{v}$ (Fig.~\ref{Fig:SpectralSupportSubSuper}e): for $90^{\circ}\!<\!\theta\!<\!\theta_{\mathrm{c}}$, $\Delta\omega^{\mathrm{max}}\!=\!\omega_{\mathrm{SPP}}-\omega_{\mathrm{o}}$ independently of $\theta$; and for $\theta_{\mathrm{c}}\!<\!\theta\!<\!180^{\circ}$,  $\Delta\omega^{\mathrm{max}}\!=\!|\widetilde{v}|k_{\mathrm{o}}'$. Here $\theta_{\mathrm{c}}$ is the spectral tilt angle corresponding to $\mathcal{P}(\theta_{\mathrm{c}})$ passing through the point $(k_{z},\tfrac{\omega}{c})\!=\!(0,\tfrac{\omega_{\mathrm{SPP}}}{c})$, such that $\tan{\theta_{\mathrm{c}}}\!=\!-\tfrac{\omega_{\mathrm{SPP}}-\omega_{\mathrm{o}}}{ck_{\mathrm{o}}'}$ (Fig.~\ref{Fig:SpectralSupportSubSuper}e). Subluminal and superluminal ST-SPPs can be distinguished in the $(k_{x},\tfrac{\omega}{c})$-plane because their spectral representations have opposite-sign curvatures. 

For \textit{any} spectral tilt angle $\theta $ smaller than the opening angle for the dielectric light-cone, the plane $\mathcal{P}(\theta)$ intersects with the SPP light-cone in two disjoint sections (Fig.~\ref{Fig:LuminalAngle}a,b). Although both sections are associated with the same $\theta$, and thus support two ST-SPPs having the \textit{same} group velocity $\widetilde{v}\!=\!c\tan{\theta}$, nevertheless the ST-SPP in the lower-frequency-section is \textit{sub}luminal, whereas the ST-SPP in the higher-frequency-section is \textit{super}luminal. This is because subluminal and superluminal values of $\widetilde{v}$ are determined with respect to that of a traditional SPP at the same frequency (Fig.~\ref{Fig:LuminalAngle}b).

\subsection{Subwavelength transverse spatial width for ST-SPPs}

By virtue of the strong spatio-temporal correlations intrinsic to the ST-SPP spectral support domain, associated with $\Delta\omega^{\mathrm{max}}$ is a maximum \textit{spatial} bandwidth $\Delta k_{x}^{\mathrm{max}}$ that determines the minimum transverse spatial width $\Delta x^{\mathrm{min}}$. In contrast to free-space ST wave packets that do \textit{not} support subwavelength $\Delta x$, we show here that ST-SPPs admit such a possibility.

\begin{figure}[t!]
  \begin{center}
  \includegraphics[width=9cm]{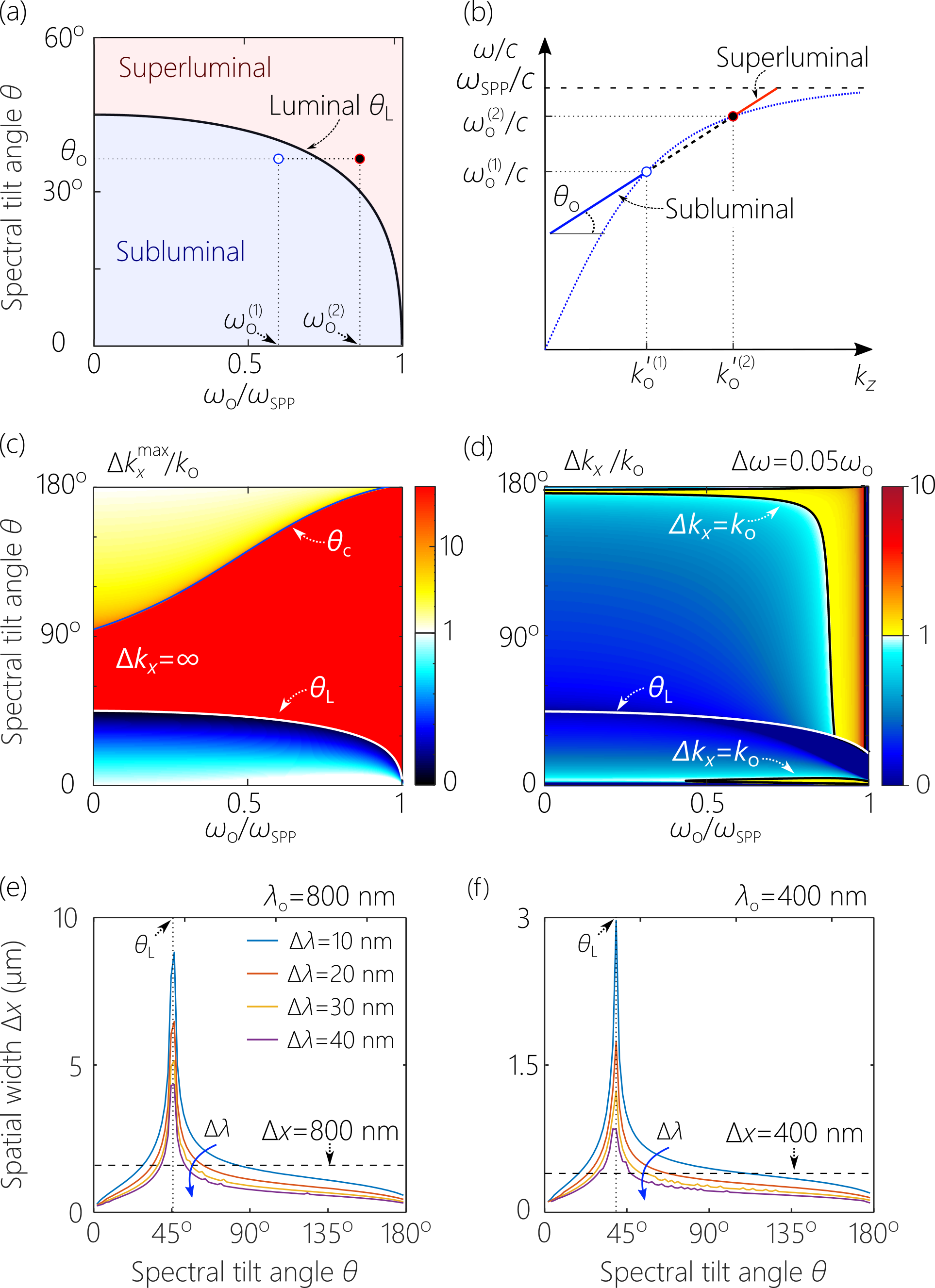}
  \end{center}
  \caption{(a) The luminal spectral tilt angle $\theta_{\mathrm{L}}$ separates the subluminal and superluminal regimes. Subluminal and superluminal ST-SPPs can share the same value of $\theta\!=\!\theta_{\mathrm{o}}$ when associated with different values of $\omega_{\mathrm{o}}$. (b) Spectral projection onto the $(k_{z},\tfrac{\omega}{c})$-plane showing a spectral plane $\mathcal{P}(\theta_{\mathrm{o}})$ intersecting with the SPP light-line in two points: lower-frequency $\omega_{\mathrm{o}}^{(1)}$ and higher-frequency $\omega_{\mathrm{o}}^{(2)}$. Below $\omega_{\mathrm{o}}^{(1)}$, $\mathcal{P}(\theta_{\mathrm{o}})$ defines a subluminal ST-SPP; above $\omega_{\mathrm{o}}^{(2)}$, $\mathcal{P}(\theta_{\mathrm{o}})$ defines a superluminal ST-SPP. (c) The normalized maximum spatial bandwidth $\Delta k_{x}^{\mathrm{max}}/k_{\mathrm{o}}$ (corresponding to the maximum temporal bandwidth $\Delta\omega^{\mathrm{max}}$) as a function of $\theta$ and $\omega_{\mathrm{o}}$. (d) Same as (c) but for a temporal bandwidth $\Delta\omega\!=\!0.05\omega_{\mathrm{o}}$. In (c) and (d), we make use of two different color scales for $\Delta k_{x}\!<\!k_{\mathrm{o}}$ (blue to white, on a \textit{linear scale}), and for $\Delta k_{x}\!>\!k_{\mathrm{o}}$ (white to red, on a \textit{logarithmic} scale). (e) Plot of $\Delta x$ against $\theta$ for different $\Delta\lambda$ when $\lambda_{\mathrm{o}}\!=\!800$~nm. Subwavelength $\Delta x$ occurs below the dashed horizontal line. (f) Same as (e) for $\lambda_{\mathrm{o}}\!=\!400$~nm.}
  \label{Fig:LuminalAngle}
\end{figure}

We obtain $\Delta k_{x}^{\mathrm{max}}$ by substituting in Eq.~\ref{Eq:kx_omega_ST-SPP} the minimum temporal frequency $\omega^{\mathrm{min}}\!=\!\omega_{\mathrm{o}}-\Delta\omega^{\mathrm{max}}$ for subluminal ST-SPPs, or the maximum temporal frequency $\omega^{\mathrm{max}}\!=\!\omega_{\mathrm{o}}+\Delta\omega^{\mathrm{max}}$ for superluminal ST-SPPs. We plot $\Delta k_{x}^{\mathrm{max}}$ in Fig.~\ref{Fig:LuminalAngle}c as a function of $\omega_{\mathrm{o}}$ and $\theta$. We normalize $\Delta k_{x}^{\mathrm{max}}$ with respect to $k_{\mathrm{o}}\!=\!\omega_{\mathrm{o}}/c$ (the free-space wave vector) to easily identify the regions in the $(\omega_{\mathrm{o}},\theta)$ parameter space that correspond to subwavelength $\Delta x^{\mathrm{min}}$, whereupon $\Delta k_{x}^{\mathrm{max}}/k_{\mathrm{o}}\!>\!1$. Two discontinuities are clear in Fig.~\ref{Fig:LuminalAngle}c: the first occurs at $\theta\!=\!\theta_{\mathrm{L}}$ (because of the mismatch in the temporal bandwidths of the upper and lower branches of luminal ST-SPPs), and the other at $\theta\!=\!\theta_{\mathrm{c}}$. There are indeed large regions in Fig.~\ref{Fig:LuminalAngle}c where $\Delta k_{x}^{\mathrm{max}}/k_{\mathrm{o}}\!>\!1$, indicating the potential for subwavelength widths for ST-SPPs. In fact, $\Delta k_{x}^{\mathrm{max}}$ is formally infinite between $\theta_{\mathrm{L}}$ and $\theta_{\mathrm{c}}$ because the SPP light-line levels off when $\omega\!\rightarrow\!\omega_{\mathrm{SPP}}$. A more realistic scenario is to consider a finite temporal bandwidth $\Delta\omega$ that is a small fraction of $\omega_{\mathrm{o}}$ as in typical pulsed laser systems. In Fig.~\ref{Fig:LuminalAngle}d we plot $\Delta k_{x}/k_{\mathrm{o}}$ for $\Delta\omega\!=\!0.05\omega_{\mathrm{o}}$ and find that whole regions in $(\omega_{\mathrm{o}},\theta)$ space retain the possibility of subwavelength transverse widths, $\Delta k_{x}/k_{\mathrm{o}}\!>\!1$. More explicitly, we plot $\Delta x$ for the ST-SPP versus $\theta$ for $\lambda_{\mathrm{o}}\!=\!2\pi c/\omega_{\mathrm{o}}\!=\!800$~nm (Fig.~\ref{Fig:LuminalAngle}e) and $\lambda_{\mathrm{o}}\!=\!400$~nm (Fig.~\ref{Fig:LuminalAngle}f), for bandwidths in the range $\Delta\lambda\!=\!10-40$~nm. Subwavelength spatial widths $\Delta x\!<\!\lambda_{\mathrm{o}}$ are realized over some range of $\theta$ even for bandwidths as small as $\Delta\lambda\!=\!10$~nm, and the useful range of $\theta$ expands for larger $\Delta\lambda$. This indicates the feasibility of utilizing available laser systems to realize plasmonic bullets with subwavelength transverse localization along $x$ on the same length scale as the SPP localization along $y$ in the dielectric.

\subsection{Spectral uncertainty and propagation length of ST-SPPs}

The model we have presented so far contains an idealization: we have assumed that each spatial frequency $k_{x}$ is associated with a \textit{single} temporal frequency $\omega$. This exact correlation results in indefinite propagation-invariance along $z$, but requires infinite energy for its realization \cite{Sezginer85JAP}. Realistic experimental resources lead to a finite `fuzziness' in the association between $k_{x}$ and $\omega$, which we term the `spectral uncertainty' $\delta\omega$ ($\delta\lambda$ on a wavelength scale); typically, $\delta\lambda\!\ll\!\Delta\lambda$ \cite{Yessenov19OE}. Each spatial frequency $k_{x}$ is therefore no longer associated with a single temporal frequency $\omega$, but rather with a narrow bandwidth $\delta\omega$ centered at the ideal $\omega$ value. We have recently shown that the propagation distance $L_{\mathrm{max}}$ of a ST wave packet is given by $L_{\mathrm{max}}\!\sim\!\tfrac{c}{\delta\omega}\,\,\tfrac{1}{|1-\cot{\theta}|}$ in free space \cite{Yessenov19OE}, which also provides an adequate estimate in other materials. In general, $L_{\mathrm{max}}$ in most cases is significantly larger than the absorption lengths of SPPs as determined by optical losses in the metal, which relaxes the need for achieving a small spectral uncertainty for ST-SPPs.

The equations for the electric and magnetic field components of the ST-SPP must be modified to accommodate this spectral uncertainty. We first replace the purely spatial spectrum $\widetilde{E}_{\mathrm{o}}(k_{x})$ with a spatio-temporal spectrum of the form $\widetilde{E}_{\mathrm{o}}(k_{x})\widetilde{h}(\Omega-\Omega(k_{x};\theta))$, where $\widetilde{h}(\Omega)$ is a narrow function of width equal to the spectral uncertainty $\delta\omega$. As such, the envelope of the $x$-component of the electric field in the dielectric becomes:

\begin{widetext}    
\begin{equation}\label{Eq:STSPPInSpaceTimeWithUncertainty}
\psi_{x}^{(\mathrm{d})}(x,y,z;t)=
\iint dk_{x}d\Omega\,\widetilde{E}_{\mathrm{o}}(k_{x})\widetilde{h}(\Omega-\Omega(k_{x};\theta))\frac{k_{x}}{k_{z}}e^{i(k_{x}x+k_{y_{\mathrm{d}}}y)}e^{-i\Omega(k_{x};\theta)(t-z/v_{\mathrm{g}})},
\end{equation}
\end{widetext}
with similar modifications to be made for the remaining field components in the dielectric and metal. These are the equations we utilized in our simulations.

\section{Simulations of the propagation of ST-SPPs}

To verify our theoretical model, we simulated the propagation of SPPs and ST-SPPs of equal initial ($z\!=\!0$) spatial and temporal widths at the interface between gold and air ($\epsilon_{\mathrm{d}}\!=\!1$). The parameters used in our simulations are provided in Table~\ref{tbl:input}, including: (1) parameters that are held equal for SPPs and ST-SPPs, which include the optical carrier wavelength $\lambda_{\mathrm{o}}$, bandwidth $\Delta\lambda$ (initial pulse width $\Delta T$), and spatial bandwidth $\Delta k_{x}$ (initial transverse spatial width $\Delta x$); (2) relevant material parameters of gold \cite{Olmon}; and (3) parameters unique to ST-SPPs, namely, the spectral tilt angle $\theta$ and the spectral uncertainty $\delta\lambda$. The widths $\Delta\lambda$, $\Delta T$, and $\Delta x$ are FHWM and $\Delta k_{x}$ is the HWHM of the relevant intensity distributions. The pulse parameters selected correspond to those for a generic mode-locked Ti:sapphire laser. Throughout, we take the intensity $I(x,y,z;t)$ to be the sum of the squared amplitudes of the field components along $x$, $y$, and $z$. Overall, the behavior of the ST-SPPs are largely unaffected by shifting the wavelength or changing the selected material parameters, and our findings are generic characteristics of all ST-SPPs. In the following we ignore the optical losses to isolate the impact of introducing spatio-temporal spectral correlations on the axial evolution of ST-SPPs, and then subsequently consider the impact of losses.  

\begin{table}
  \caption{SPP and ST-SPP Simulation Parameters.}
  \label{tbl:input}
  \begin{tabular}{llll}
    \hline
    Parameter & Symbol & Value & Units\\
    \hline
    carrier wavelength & $\lambda_{\mathrm{o}}$ & 800 & nm\\
    spectral bandwidth & $\Delta\lambda$ & 10 & nm\\
    pulse width & $\Delta T$ & 94 & fs\\
    spatial bandwidth & $\Delta k_{x}$ & 0.586 & rad/$\mu$m\\
    transverse spatial width & $\Delta x$ & 3.2 & $\mu$m\\
    \\
    plasma frequency \cite{Olmon} & $\omega_{\mathrm{p}}$ & 12.84 & rad*PHz\\
    damping rate \cite{Olmon} & $\Gamma$ & 72.93     & rad*THz\\
    dielectric constant & $\epsilon_{\mathrm{d}}$ & 1 & ---\\
    \\
    spectral uncertainty & $\delta\lambda$ &10 &pm\\
    spectral tilt angle & $\theta$ & 38 & $^{\circ}$\\
    \hline
\end{tabular}

\end{table}

\subsection{Propagation of traditional SPP wave packets}

\begin{figure*}
\centering
\begin{center}
\includegraphics[width=16.4cm]{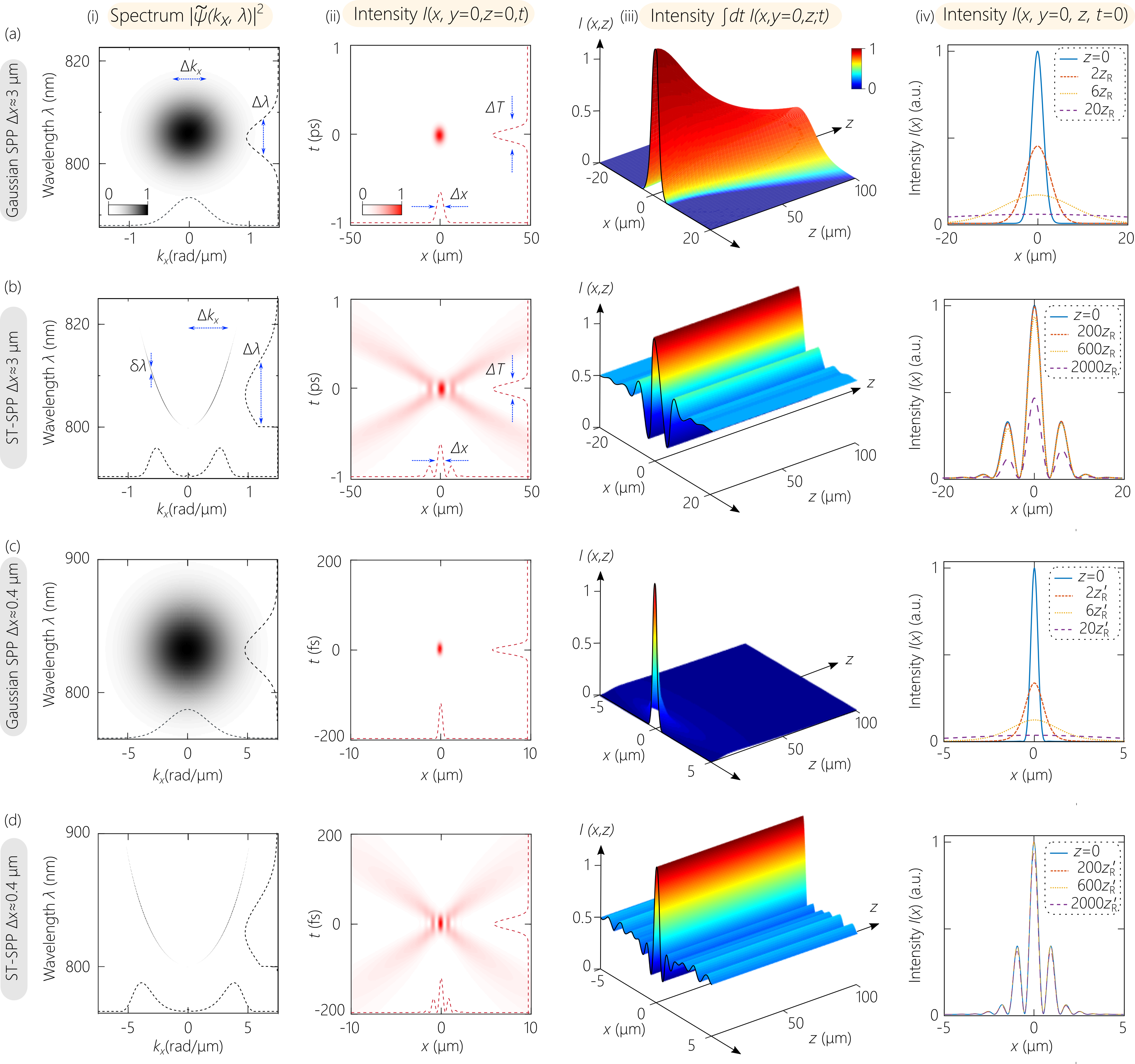}
\end{center}
\caption{Axial dynamics of traditional SPPs and ST-SPPs (all at $\lambda_{\mathrm{o}}\!=\!800$~nm). (a) A traditional SPP with Gaussian spatial and temporal spectra such that $\Delta x\!\approx\!3$~$\mu$m. (b) A ST-SPP with $\Delta\!\approx\!3$~$\mu$m. (c) A traditional SPP with Gaussian spatial and temporal spectra such that $\Delta x\!\approx\!400$~nm. (d) A ST-SPP with $\Delta x\!\approx\!400$~nm. Column (i) shows the spatio-temporal spectral intensity, $I(x,y\!=\!0,z\!=\!0;t)$; column (ii) shows the intensity profile $I(x,y\!=\!0,z\!=\!0;t)$; column (iii) shows the time-averaged intensity $I(x,z)\!=\!\int\!dt\,I(x,y\!=\!0,x;t)$; and column (iv) shows sections through (iii) at different axial positions. The Rayleigh range in (a) and (b) is $z_{\mathrm{R}}\!=\!25$~$\mu$m, and in (c) and (d) $z_{\mathrm{R}}'\!=\!700$~nm. In column (i) we plot the projected marginal distributions for the spatial and temporal spectra.}\label{Fig:DiffractionSPP}
\end{figure*}

We assume for the traditional SPP a generic Gaussian spatio-temporal spectrum that is separable with respect to the spatial and temporal frequencies, $\widetilde{\psi}(k_{x},\Omega)\!\propto\!\exp{\left(-2\ln{2}\tfrac{k_{x}^{2}}{(\Delta k_{x})^{2}}\right)}\exp{\left(-2\ln{2}\tfrac{\Omega^{2}}{(\Delta\Omega)^{2}}\right)}$, where $\Delta\Omega$ and $\Delta k_{x}$ are the FWHM of the spatial and temporal spectra (Table~\ref{tbl:input}), as depicted in Fig.~\ref{Fig:DiffractionSPP}a-i. This form of $\widetilde{\psi}(k_{x},\Omega)$ results in a wave packet that separates at $z\!=\!0$ into a product of a Gaussian \textit{pulse} of width $\Delta T\!\approx\!94$~fs and a Gaussian transverse \textit{spatial} profile of width $\Delta x\!\approx\!3$~$\mu$m (Fig.~\ref{Fig:DiffractionSPP}a-ii). The widths $\Delta T$ and $\Delta x$ are both FWHM of the intensity profile $I(x,y\!=\!0,z\!=\!0;t)$. Propagation along the gold surface leads to diffraction along $x$ with a Rayleigh range $z_{\mathrm{R}}\!=\!25$~$\mu$m, and dispersive broadening is negligible here over multiple Rayleigh ranges (see below); see Fig.~\ref{Fig:DiffractionSPP}a-iii where we plot the axial dynamics of the time-averaged intensity $I(x,z)\!=\!\int\!dt\;I(x,y\!=\!0,z;t)$. To highlight the diffractive broadening of the spatial profile, we plot in Fig.~\ref{Fig:DiffractionSPP}-iv sections through $I(x,z)$ comparing the initial intensity profile at $z\!=\!0$ with those at $z\!=\!2z_{\mathrm{R}}$, $6z_{\mathrm{R}}$, and $20z_{\mathrm{R}}$.
$|\widetilde{\psi}(k_{x},\lambda)|^{2}$

\subsection{Propagation of ST-SPP wave packets}

To facilitate comparison with the preceding traditional Gaussian SPP wave packet, we design the spatio-temporal spectrum of the ST-SPP (Fig.~\ref{Fig:DiffractionSPP}b-i) such that the spatial width at the pulse center is $\Delta x\!\approx\!3$~$\mu$m, and the pulse width at $x\!=\!0$ is $\Delta T\!\approx\!94$~fs. We calculate the ideal spectral trajectory $\Omega(k_{x};\theta)$ from Eq.~\ref{Eq:kx_omega_ST-SPP} and then add the finite spectral uncertainty, which is modelled as a Gaussian distribution with FWHM $\delta\lambda\!\ll\!\Delta\lambda$ that we convolve with $\Omega(k_{x};\theta)$. To realize the target values for $\Delta x$ and $\Delta T$ for the ST-SPP, we take the projected temporal spectrum to be a Gaussian distribution that is truncated at $\lambda_{\mathrm{o}}\!=\!800$~nm (in the vicinity of $k_{x}\!=\!0$), select $\theta\!=\!38^{\circ}$, and reduce the amplitude of the spatial spectrum in the vicinity of $k_{x}\!=\!0$ to avoid a van-Hove-like singularity at the spectral edge \cite{Gilmore04QM1D} (Fig.~\ref{Fig:DiffractionSPP}b-i).

The spatio-temporal spectrum projected onto the $(k_{x},\lambda)$-plane (as shown in Fig.~\ref{STSPP_concept}b) resembles in its basic features the spectral structures utilized in previous experimental demonstrations of free-space ST wave packets \cite{Kondakci17NP,Yessenov19PRA}. We plot in Fig.~\ref{Fig:DiffractionSPP}b-ii the spatio-temporal intensity profile at $z\!=\!0$, where the characteristic X-shaped profile associated with ST wave packets is clear (as shown also in Fig.~\ref{STSPP_concept}b). The profile consists of an intense central peak at $x\!=\!0$ and $t\!=\!0$ from which decaying tails extend. Consequently, the time-averaged intensity in Fig.~\ref{Fig:DiffractionSPP}b-iii comprises a narrow central peak atop a broad background representing the integration of the intensity tails. Over an axial distance of 100~$\mu$m ($4z_{\mathrm{R}}$ for the preceding Gaussian SPP), the ST-SPP has undergone no change. Indeed, the axial intensity drops only at much larger distances (Fig.~\ref{Fig:DiffractionSPP}b-iv) that far exceed the plasmon decay length, so that we can consider the ST-SPP to be propagation-invariant for all practical purposes.

\subsection{Propagation of subwavelength-width ST-SPPs}

As discussed above, ST-SPPs can be realized with subwavelength $\Delta x$. Indeed, because the SPP light-cone extends below the dielectric light-line, even a traditional monochromatic SPP can be produced with subwavelength $\Delta x$, but such a SPP will diffract very rapidly (e.g., the effective Rayleigh range when $\Delta x\!=\!400$~nm is $z_{\mathrm{R}}'\!=\!500$~nm) -- whereas the ST-SPP will not. We verify this in the simulations plotted in Fig.~\ref{Fig:DiffractionSPP}c,d where we compare the axial dynamics for a Gaussian SPP (Fig.~\ref{Fig:DiffractionSPP}c) with a ST-SPP (Fig.~\ref{Fig:DiffractionSPP}d) of equal widths $\Delta x\!=\!400$~nm at $\lambda_{\mathrm{o}}\!=\!800$~nm. The parameters of this simulations are as follows: $\Delta\lambda\!=\!50$~nm, $\Delta T\!=\!36$~fs, $\theta\!=\!15^{\circ}$, and $\Delta k_{x}\!\approx\!4.1$~rad/$\mu$m (the same $\Delta x$ for the ST-SPP can be achieved with larger $\Delta T$ by decreasing $\theta$). The initial profile of the Gaussian SPP broadens after propagating a few microns (Fig.~\ref{Fig:DiffractionSPP}c-iii,iv), whereas the ST-SPP (Fig.~\ref{Fig:DiffractionSPP}d-i,ii) maintains its profile over a much larger propagation distance (Fig.~\ref{Fig:DiffractionSPP}d-iii,iv). Indeed, the propagation distance is dictated by the spectral uncertainty $\delta\lambda$ and $\theta$, almost independently of $\Delta x$ \cite{Yessenov19OE}. Therefore, in light of the typical plasmon decay lengths, the ST-SPP can be considered propagation-invariant for \textit{any} $\Delta x$, even for subwavelength widths.

\begin{figure*}
    \centering
    \includegraphics[width=16.6cm] {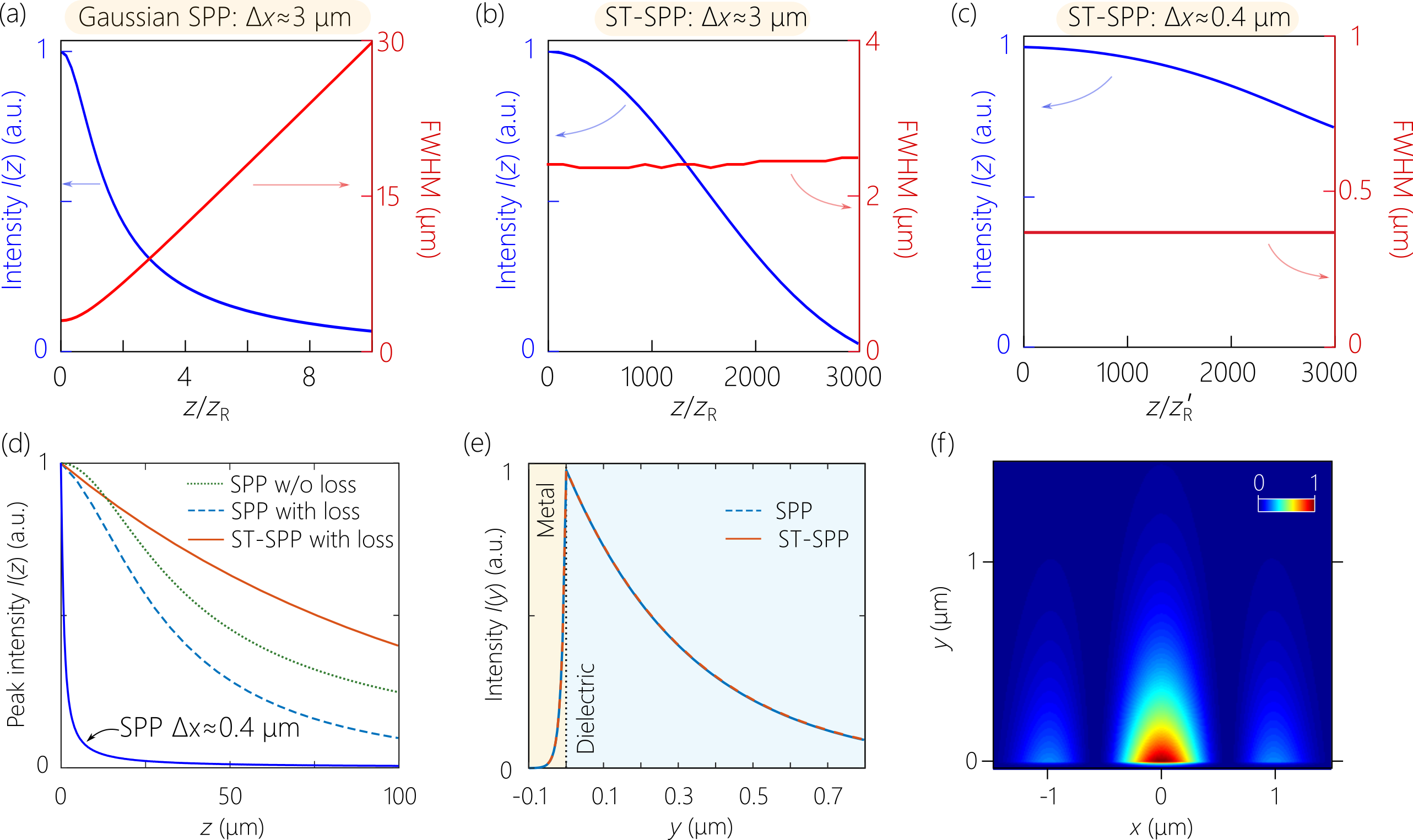}
    \caption{(a) Evolution of the on-axis time-averaged intensity $I(0,z)$ for the SPP with $\Delta x\!\approx\!3$~$\mu$m (Fig.~\ref{Fig:DiffractionSPP}a); (b) for a ST-SPP with $\Delta x\!\approx\!3$~$\mu$m (Fig.~\ref{Fig:DiffractionSPP}b); and (c) for a ST-SPP with $\Delta x\!\approx\!400$~nm (Fig.~\ref{Fig:DiffractionSPP}c). (d) Axial evolution of $I(0,z)$ for SPPs and ST-SPPs of widths $\Delta x\!\approx\!3$~$\mu$m and $\Delta x\!\approx\!400$~nm in presence and absence of plasmonic losses. For the SPP with $\Delta x\!\approx\!400$~nm, the plots in presence and absence of losses coincide. For ST-SPPs, there is no observable decay in absence of losses over the plotted length scale, and in presence of losses the two curves for $\Delta x\!=\!3$~$\mu$m and $\Delta x\!=\!400$~nm coincide. (e) Plots of the plasmon localization along $y$, $I(x\!=\!0,y,z\!=\!0;t\!=\!0)$, for a SPP and a ST-SPP with $\Delta x\!\approx\!3$~$\mu$m. (f) Transverse intensity profile $I(x,y,z\!=\!0;t\!=\!0)$ when $\Delta x\!=\!400$~nm.}
    \label{fig:losses_compare}
\end{figure*}

We summarize the axial dynamics of the on-axis intensity and spatial width $\Delta x$ in Fig.~\ref{fig:losses_compare}a-c. Figure~\ref{fig:losses_compare}a displays the results for the Gaussian SPP of initial width $\Delta x\!\approx\!3$~$\mu$m, which behaves in step with a Gaussian beam in free space. The peak intensity of the ST-SPP with $\Delta x\!\approx\!3$~$\mu$m decays over distances that exceed those for the Gaussian SPP by a factor of $\approx\!300$ (reaching 1/e of its initial value until $z\!=\!2100z_{\mathrm{R}}$; Fig.~\ref{fig:losses_compare}b). Surprisingly, the width $\Delta x$ does \textit{not} increase accordingly. Instead, energy leaks into the tails of the X-shaped spatio-temporal profile (thereby contributing to the background in the time-averaged intensity), while leaving the central peak almost unchanged. Similar behavior is observed for the ST-SPP with $\Delta x\!=\!400$~nm (except that $z_{\mathrm{R}}'\!<\!z_{\mathrm{R}}$; Fig.~\ref{fig:losses_compare}c). 

\subsection{Effect of SPP losses and dispersion}

We now examine the effect of optical losses inherent to SPPs due to the dissipative electronic dynamics in the metal by utilizing the imaginary part of the metal permittivity in the Drude model, $\text{Im}\left\{\epsilon_{\mathrm{m}}(\omega)\right\}=\frac{\omega_{\mathrm{p}}^{2}\Gamma}{\omega(\omega^{2}+\Gamma^{2})}$. The narrow bandwidth considered here ($\Delta\lambda\!=\!10$~nm) allows us to ignore the frequency dependence of $\epsilon_m$, so that the intensity absorption coefficient is $\alpha\!\approx\!84$~$\mu$m$^{-1}$. The general features of the previous results are unaffected except for an overall drop in intensity with propagation distance. Two phenomena contribute to the decay of the on-axis intensity of a SPP: absorption and diffractive spreading -- whereas the corresponding decay of a ST-SPP results \textit{solely} from absorption. The decay length for a SPP due to diffraction drops quadratically with $\Delta x$, and can indeed become smaller than the absorption length.

We plot in Fig.~\ref{fig:losses_compare}d the on-axis intensity for the Gaussian SPP and the ST-SPP with $\Delta x\!\approx\!3$~$\mu$m in presence and absence of losses (the intensity remains essentially $100\%$ for the ST-SPP in absence of losses). At this $\Delta x$, the decay length due to diffraction is smaller that that due to absorption. Consequently, the axial decay rate of the ST-SPP in \textit{presence} of losses is lower than that of the traditional SPP of equal $\Delta x$ in \textit{absence} of losses (of course, the decay rate increases further for the SPP in presence of both diffraction and losses). This point is further emphasized by considering the subwavelngth width of $\Delta x\!=\!400$~nm: the SPP decays very rapidly due to diffractive spreading, and optical losses therefore have virtually no effect; whereas the decay rate of the subwavelength ST-SPP in presence of losses coincides with the ST-SPP having $\Delta x\!\approx\!3$~$\mu$m. Therefore, ST-SPPs have an unsurpassed advantage in presence of losses with respect to traditional SPPs as $\Delta x$ decreases, and dramatically so in the subwavelength domain.

Traditional SPPs are subject to intrinsic GVD due to the curvature of the SPP light-cone, which is further affected by the wavelength-dependence of $\epsilon_{\mathrm{m}}$ and $\epsilon_{\mathrm{d}}$. The dispersion length $L_{\mathrm{D}}$ of the SPP is determined by $\epsilon_{\mathrm{d}}$, $\epsilon_{\mathrm{m}}$, $\lambda_{\mathrm{o}}$, and the bandwidth $\Delta\lambda$. For a pulse width of 100-fs at $\lambda_{\mathrm{o}}\!=\!800$~nm at a gold-air interface, $L_{\mathrm{D}}\!\sim\!6$~cm. For the subwavelength SPP with $\Delta x\!=\!400$~nm (Fig.~\ref{Fig:DiffractionSPP}c), the bandwidth is $\Delta\lambda\!=\!50$~nm and $L_{\mathrm{D}}\!\sim\!2.4$~mm. In contrast, in all cases, the ST-SPP is immune to GVD.

The confinement of the ST-SPP along $x$ does not affect its localization along $y$. Indeed, the evanescent ST-SPP fields in the metal and dielectric are indistinguishable from those of a SPP having the same initial spatial and temporal widths (Fig.~\ref{fig:losses_compare}e). Finally, localization within the dielectric can be matched by localization along $x$ in the subwavelength regime, as is clear from the transverse intensity profile shown in Fig.~\ref{fig:losses_compare}f, thereby confirming the potential for realizing a propagation-invariant SPP that is localized in all dimensions.

\section{Discussion}

\subsection{Preparation of ST-SPPs}

One potential approach to launching ST-SPPs is to start from free-space ST wave packets prepared in the form of a light sheet, and then couple the optical field to the metal-dielectric interface via the standard Kretschmann configuration \cite{Kretschman71ZP,Raether88Book}. The ST pulsed light sheet can be prepared by the spatio-temporal spectral-phase modulation strategy we recently introduced \cite{Kondakci17NP,Yessenov19OPN}. Briefly, a diffraction grating resolves the spectrum of a plane-wave pulse in space, and a spatial light modulator impresses upon the wave front a two-dimensional phase distribution in which each wavelength is assigned a prescribed spatial frequency $\Omega(k_{x};\theta)$ in accordance with Eq.~\ref{Eq:kx_omega_ST-SPP} -- which differs from the corresponding free-space relationship in which $\Omega(k_{x};\theta)$ is a conic section. The modulated spectrum is then reconstituted by a grating, thereby producing the ST wave packet in the form of a light sheet that is localized along $x$ and uniform along $y$. Because the temporal frequencies $\omega$ and the spatial frequencies $k_{x}$ are invariant across interfaces parallel to the $x$-axis, this ST light sheet can couple to a SPP. Reaching ultrabroad temporal bandwidths (ultrashort pulses) and strong transverse confinement (large spatial bandwidth) runs up against the technical limitations of spatial light modulators. However, lithographically inscribed phase plates can be used to circumvent these restrictions \cite{Kondakci18OE}. This procedure can produce ST-SPPs of spatial widths of a few microns \cite{Kondakci17NP}, and new approaches are needed for realizing subwavelength ST-SPPs.

\begin{figure*}[t!]
    \includegraphics[width=16.6cm] {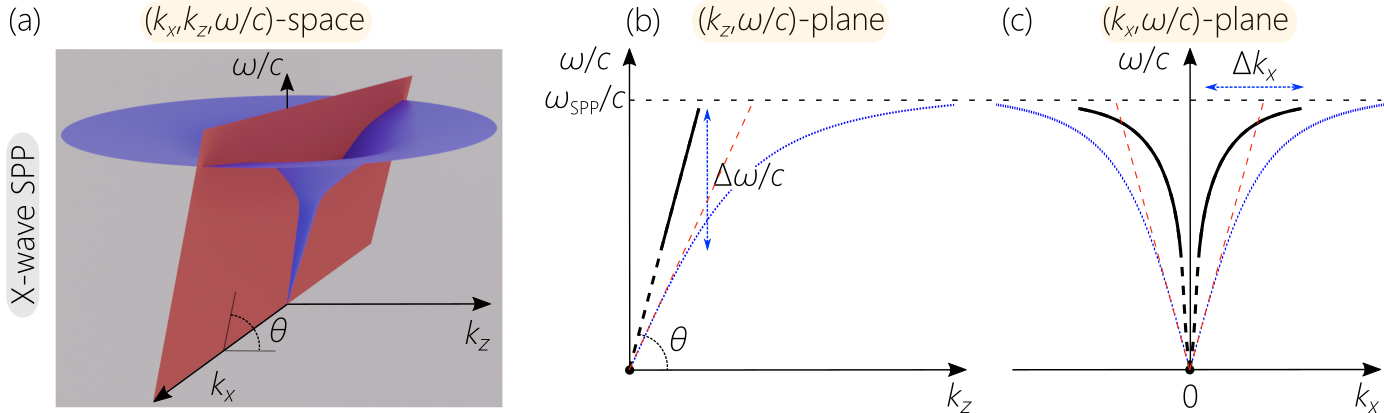}
    \caption{X-wave ST-SPPs. (a) The SPP light-cone intersecting with a tilted spectral plane $\mathcal{P}(\theta)$ passing through the origin. (b) Spectral projection of the X-wave ST-SPP onto the $(k_{z},\tfrac{\omega}{c})$-plane and (c) onto the $(k_{x},\tfrac{\omega}{c})$-plane. The thin blue dotted curves are the SPP light-lines, the thin dashed red lines are the dielectric light-lines, and the thick black lines are the projected spectral support domains of the X-wave ST-SPP.}
    \label{Fig:XWwave}
\end{figure*}

\subsection{Other avenues of research}

Other plasmonic configurations may be envisioned in which longer propagation distances are possible, which then stand to benefit from the propagation-invariance of ST-SPPs more so than the metal-dielectric-interface scenario studied here, where attenuation in the metal dominates, such as dielectric-metal-dielectric systems where the propagation length can be increased by an order of magnitude \cite{Berini2009} and metal-dielectric-metal configurations \cite{Dastmalchi16AOM}. A unique feature of ST-SPPs is the ready access to negative group velocities \cite{Feigenbaum09OE} at the metal-dielectric interface, which paves the way to convenient realization of backward phase-matching \cite{Lan15NM} by spatio-temporally structuring the optical field rather than relying on a structured plasmonic platform.

It has been recently elucidated that the enhancement in propagation distance of a ST wave packet over that of a traditional wave packet is determined by the `degree of classical entanglement' \cite{Qian11OL,Kagalwala13NP,Berg15Optica,Aiello15NJP,Kondakci19OL} of the field's spatio-temporal spectrum as determined by the spectral uncertainty $\delta\lambda$ \cite{Yessenov19OE,Kondakci19OL}. Because long propagation distances are not required in plasmonic settings, we can increase $\delta\lambda$, which alleviates the strict requirements in spatio-temporal spectral synthesis while reducing the background in the time-averaged intensity, thus enhancing the field transverse localization.

Another possibility is to create incoherent \cite{Norrman16EPL,Chen18PRA} ST-SPPs using LEDs (light-emitting diodes) or superluminescent diodes, both of which have been exploited recently in producing incoherent free-space ST wave packets \cite{Yessenov19Optica,Yessenov19OL}. This may help reduce the complexity of laser-based SPP sensing systems. Yet another opportunity is to utilize the omni-resonance of certain broadband ST wave packets in which their full bandwidth resonates with narrow-linewidth cavity structures \cite{Shiri19arxiv}. Moreover, we have recently verified a host of unexpected phenomena regarding changes in the group velocity that occur upon the refraction of ST wave packets at planar interfaces between two dielectrics (such as group-velocity invariance and group-delay cancellation), which can be put to the service of ST-SPPs \cite{Bhaduri19unpublished}.

Finally, we investigated here a family of ST-SPPs at a metal-dielectric interface resulting from restricting the spectral support of SPPs to the intersection of the SPP light-cone with the tilted spectral plane given by $\Omega\!=\!(k_{z}-k_{\mathrm{o}}')c\tan{\theta}$. Other families of ST-SPPs can be envisioned, such as X-wave SPPs where the spectral plane passes through the origin $(k_{x},k_{z},\tfrac{\omega}{c})\!=\!(0,0,0)$ and is thus given by $\omega\!=\!k_{z}c\tan{\theta}$ \cite{Lu92IEEEa,Saari97PRL}. In this case, the group velocities are always superluminal and positive-valued, see Fig.~\ref{Fig:XWwave}. Other possibilities include SPP analogs of Brittingham's focus-wave modes \cite{Brittingham83JAP,Reivelt00JOSAA,Reivelt02PRE}. The realization of SPPs in these two families require synthesizing 1D versions of them, which was achieved recently \cite{Yessenov19PRA}. Indeed, the full classification of free-space ST light sheets \cite{Yessenov19PRA} can have corresponding ST-SPP analogs.

\section{Conclusions}

We have presented a novel \textit{propagation-invariant surface wave packet that is localized in all dimensions}: space-time surface plasmon polaritons that travel in a straight line at a metal-dielectric interface. The wave packet is localized in the direction normal to the interface by the usual SPP confinement; is localized parallel to the interface by ST confinement that prevents diffractive spreading no matter how narrow the SPP width; and is localized axially by means of the pulse width, and ST confinement prevents dispersive spreading. Although we made use of the Drude modal for the metal permittivity, the results are wholly independent of the specifics of this model. Adopting a different model modifies the surface of the SPP light-cone, which then provides a new spectral support domain for the ST-SPP at the intersection with the tilted spectral plane.

We have verified through analysis and numerical simulations the propagation-invariance of ST-SPPs over distances relevant for SPPs, which can extend to other plasmonic configurations in which longer propagation distances are accessible. Moreover, we demonstrated that the group velocity of the ST-SPP can be readily controlled independently of the material properties to realize subluminal, superluminal, and negative values by varying its spatio-temporal spectral structure. Furthermore, we showed that the ST-SPP can be localized to subwavelength dimensions in the transverse spatial dimension -- on the same scale as the localized evanescent plasmon field in the dielectric -- without affecting access to arbitrary group velocities or extended propagation distances. At the heart of this new class of surface waves is a fundamental principle: by introducing tight spatio-temporal correlations into the wave packet spectrum, unprecedented control over its propagation characteristics at the surface can be achieved, including tunable group velocities in the subluminal, superluminal, and even negative-velocity regimes. These results may find applications in new nonlinear plasmonic phase-matching configurations that benefit from the additional field enhancement resulting from localization in all dimensions, and the controllable group velocity at the interface.


\section*{Acknowledgments}
This work was supported by the U.S. Office of Naval Research (ONR) under contract N00014-17-1-2458.

\bibliography{diffraction}

\end{document}